# Structural transitions in hybrid improper ferroelectric $Ca_3Ti_2O_7$ tuned by site-selective iso-valent substitutions: a first-principles study

C. F. Li,[1] S. H. Zheng,[1] H. W. Wang,[1] J. J. Gong,[1] X. Li,[1] Y. Zhang,[1] K. L. Yang,[1] L. Lin,[1] Z. B. Yan,[1] S. Dong,[2,*] and J. –M. Liu[1,3]

[1]*Laboratory of Solid State Microstructures and Innovative Center of Advanced Microstructures, Nanjing University, Nanjing 210093, China*

[2]*School of Physics, Southeast University, Nanjing 211189, China*

[3]*Institute for Advanced Materials, South China Normal University, Guangzhou 510006, China*

**[Abstract]** $Ca_3Ti_2O_7$ is an experimentally confirmed hybrid improper ferroelectric material, in which the electric polarization is induced by a combination of the coherent $TiO_6$ octahedral rotation and tilting. In this work, we investigate the tuning of ferroelectricity of $Ca_3Ti_2O_7$ using iso-valent substitutions on Ca-sites. Due to the size mismatch, larger/smaller alkaline earths prefer *A'*/*A* sites respectively, allowing the possibility for site-selective substitutions. Without extra carriers, such site-selected iso-valent substitutions can significantly tune the $TiO_6$ octahedral rotation and tilting, and thus change the structure and polarization. Using the first-principles calculations, our study reveals that three substituted cases (Sr, Mg, Sr+Mg) show divergent physical behaviors. In particular, $(CaTiO_3)_2SrO$ becomes non-polar, which can reasonably explain the suppression of polarization upon Sr substitution observed in experiment. In contrast, the polarization in $(MgTiO_3)_2CaO$ is almost doubled upon substitutions, while the estimated coercivity for ferroelectric switching does not change. The $(MgTiO_3)_2SrO$ remains polar but its structural space group changes, with moderate increased polarization and possible different ferroelectric switching paths. Our study reveals the subtle ferroelectricity in the $A_3Ti_2O_7$ family and suggests one more practical route to tune hybrid improper ferroelectricity, in addition to the strain effect.

* Corresponding author. Email: sdong@seu.edu.cn

## I. Introduction

A ferroelectric is a material with a spontaneous electric polarization ($P$), which can be reversed by electric field ($E$). Ferroelectric materials have been widely used in capacitors, piezoelectric sensors/transducers, random access memories, etc [1-3]. More ferroelectrics with superior performance are still in appealing, triggering continuous searching for various properties and functionalities [4].

Conventionally, ferroelectrics can be roughly classified into two categories in term of polarization generation mechanisms. One class falls in the displacive-type category where spontaneous polarization is caused by the displacement of atoms related to their high symmetric positions in paraelectric phase, which is more frequently seen in inorganic materials. The other class belongs to the ordering of polar group (or collective proton transfer), which is more common in organic materials [5]. For the first class, traditionally, the polar phonon modes are usually responsible for the ferroelectric structural transition, i.e. ion displacements. However, some novel mechanisms for ferroelectricity generation have been found in recent years, enriching the quantum physics of ferroelectricity and potentials of additional functionalities. For examples, a number of magnetism-induced ferroelectrics (the so-called type-II multiferroics) have been found, and the magnetoelectric coupling and inter-control functionalities are being realized [6, 7]. In these improper ferroelectrics, the polar phonon modes are no longer the driving force of ferroelectricity.

The scope of improper ferroelectrics can be even broader. One highly promising branch is the so-called hybrid improper ferroelectrics [8, 9], as first predicted in even-layered Ruddlesden-Popper (RP) perovskite oxides, e.g. $Ca_3Mn_2O_7$ (CMO) and $Ca_3Ti_2O_7$ (CTO). In the RP series, the layered structure makes them attractive as natural atomic-scale superlattices. The origin of their polarizations stems from a combination of coherent oxygen octahedral rotation mode and tilting mode, neither of which is polar (see Fig. 1). Furthermore, the octahedral tilting mode can induce weak ferromagnetism in CMO due to spin canting, while the octahedral rotation mode triggers the magnetoelectricity [8]. These characteristics of hybrid improper ferroelectricity make the 327-RP perovskite highly interesting. Experimentally, the hybrid improper ferroelectricity has been verified in $(Ca,Sr)_3Ti_2O_7$ [10] and $[1-x](Ca_ySr_{1-y})_{1.15}Tb_{1.85}Fe_2O_7$-$[x]$CTO [11], while the CMO is more complicated beyond the original prediction [12].

Further theoretical studies found multiple rotation/tilting modes in these 327-RP series beyond the ground state one, which could lead to proximate energies but different space groups. Some of these space groups are nonpolar, while some of them are still polar. Thus, it is scientifically interesting to investigate the transitions between these structures, which can play as the intermediate states of ferroelectric switching or new ground states upon perturbations. For example, Nowadnick and Fennie identified possible ferroelectric switching pathways in CTO, which were via an orthorhombic twin domain or via an antipolar structure [13]. These switching pathways are relevant to ferroelectric domain walls and coercivity. A small coercivity is crucial for energy saving in applications of ferroelectricity. In addition, Lu and Rondinelli predicted the polar-to-nonpolar transitions achieved in CTO and magnetic Ti-site substituted CTO film via epitaxial strain [14, 15] .

Besides epitaxial strain, iso-valent substitution is also a very powerful tool to fine tune the structure, and then the hybrid improper ferroelectricity. The subtle balance between the rotation/tilting modes can be easily infected by the substitution, although no extra carriers are introduced into the system. Experimentally, Sr-substituted CTO has been synthesized, whose ferroelectric polarization is significantly suppressed comparing with that of CTO [10], although an early theoretical calculation recommended the enhancement of polarization in general by larger size ion substitution [16].

In this work, the structure and ferroelectricity of iso-valent substituted CTO is studied using the density functional theory (DFT) calculations. First, the Perdew-Burke-Ernzerhof revised for solid (PBEsol) parameterization of the generalized gradient approximation (GGA) plus on-site Coulomb interaction $U$ method is employed to re-calculate the electric polarization of CTO, intending to give a more precise consistency with measured values, since previous DFT studies overestimate the polarization [10, 13, 14]. Second, based on the accurate CTO result, the site-selective substitutions are studied to tune the structure as well as the ferroelectricity. For the RP structure (see Fig. 1), the chemical formula of CTO can be re-written as $(A\text{TiO}_3)_2A'\text{O}$ with $A$=Ca(1) and $A'$=Ca(2). There are two inequivalent Ca sites, which lead to the possibility for the site-selective substitution. Here a larger size ion Sr and a smaller size ion Mg are considered in the iso-valent substitutions, leading to three compounds: $(\text{CaTiO}_3)_2\text{SrO}$, $(\text{MgTiO}_3)_2\text{CaO}$, and $(\text{MgTiO}_3)_2\text{SrO}$. The reason for such configurations will be explained in the following sections. Our DFT calculations on these three compounds give interesting and divergent results.

## II. Model systems and computational methods

### 2.1. Structural consideration of model systems

The 327-type CTO has its ferroelectric structure (space group $A2_1am$) and hypothetical paraelectric structure (space group *Amam*), as shown in Fig. 1(d) and 1(b) respectively. The structures are made up of a perovskite (P) block in which the Ca(2) ion locates at the body center (*A'*-site), and a rock-salt (R) block consisting of arranged Ca(1) ions at the *A*-sites, and the O(3) ions in between the adjacent P blocks. The P block is built up of two layers of corner-sharing TiO$_6$ octahedra along the *c*-axis. The *A*-cation is at the 9-coordinate site and the *A'*-cation at the 12-coordinate site. The different coordination numbers of *A*-sites in the R blocks and *A'*-sites in the P blocks lead to the preferential substitution of large/small ions in the P/R blocks due to the large/small space. This preference was suggested and supported by experiments [10, 11]. This preference will be further numerically checked in our following DFT calculations.

Taking the highly symmetric *I*4/*mmm* structure as the starting point (Fig. 1(a)), the octahedral tilting/rotation alone, generates the *Amam*/*Acam* structure respectively, neither of which breaks the space-inversion symmetry, i.e. nonpolar. In particular, the *Amam* phase (Fig. 1(b)) is obtained after the octahedral $a^-a^-c^0$ tilting with respect to the tetragonal *I*4/*mmm* structure. The ferroelectric ground state of CTO is *A*2$_1$*am* phase (Fig. 1(d)), in which the TiO$_6$ octahedra rotate alternately in the *ab*-plane around the *c*-axis, represented as $a^0a^0c^+$ in the Glazer notation with respect to the *Amam* parent phase. The simultaneous presence of the tilting and rotation can lead to the upward Ca(1) displacements and downward Ca(2) displacements along the *a*-axis with reference to the Ti ions, as shown in Fig. 1(d), where the arrows indicate the ionic displacements. In correspondence, the O(3) ions also coherently shift downward and the O(2) ions upward (not shown by arrows). Consequently, a separation of the charge centers of anions and cations gives rise to a net *P*.

According to the schematic diagrams for the octahedral tilts/rotations [17], there are many other possibilities for the combinations of octahedral rotations/tiltings. Thus the structures and ferroelectricity of 327-RP series can be even more complicated. Thus, although in Ref. [16], the strategy of isovalent substitution at *A*/*A*' site was proposed for (CaSnO$_3$)$_2$CaO, it remains necessary to carefully check the iso-valent substitution in (CaTiO$_3$)$_2$CaO, especially considering the disagreement between experimental result [10] and DFT prediction [16].

*2.2. DFT calculations*

The DFT calculations are done based on the projected augmented wave (PAW) pseudo-potentials as implemented in Vienna *ab initio* Simulation Package (VASP) [18-20]. The electron interactions are described using PBEsol [21] parametrization of GGA [22]. The plane-wave cutoff is set to 600 eV. The Monkhorst-Pack *k*-point mesh is 8×8×2. The polarization is calculated using the Berry phase approach [23]. To analyze the layer-dependent layer-by-layer dipole moments, the Born effective charge (BEC) model is also used, and the effective charge for each ion is calculated via density functional perturbation theory (DFPT) [24].

In addition, the Dudarev implementation [25] is adopted to add an on-site Coulomb interaction $U_{eff}$ (=$U$-$J$) to the 3$d$ orbitals of Ti. Nominally, the 3$d$ orbitals of Ti$^{4+}$ are empty, rendering a band insulator instead of a Mott insulator for CTO. However, the including of $U$ in DFT calculation is practically necessary as the self-interaction correction or PBE overestimation of covalency. In fact, the first DFT calculation based on local spin density approximation (LSDA)+$U$ overestimated ferroelectric $P$ for both CMO (~5.0 μC/cm$^2$ in DFT but not experimentally measured yet) and CTO (~20 μC/cm$^2$ in DFT, ~8 μC/cm$^2$ in measurement of single crystals) [8]. The choice of PBEsol can improve the precision of calculated structures (and thus the polarization ~17 μC/cm$^2$) [14], but is still overestimated. To pursuit a better match between DFT results and experimental values, the PBEsol+$U$ would be helpful, which is essential for reliable prediction considering the subtle structure transitions in the RP series.

For each compound, one unit cell contains four formula units (f.u.) with total 48 atoms, in either the paraelectric or ferroelectric phase. For CTO, the experimentally determined atomic positions and lattice constants in the $A2_1am$ space group are adopted as the initial structure of the ferroelectric state. Then the atomic positions and lattice constants are fully relaxed iteratively until the Hellman-Feynman forces on every atoms are converged to be less than 3 meV/Å. For the other three compounds, we also fully relax the structures of $A2_1am$ and other possible space groups to find the ground states.

## III. Results and discussion

### 3.1. Choice of Hubbard $U_{eff}$

We first check the effect of Hubbard $U_{eff}$ using the CTO as the object whose experimental data are plenty for comparison. Then the optimal $U_{eff}$ value will be used for the following calculations of the substituted compounds. Given a set of $U_{eff}$ ranging from 0 to 6 eV stepped by 1 eV, the ferroelectric state ($A2_1am$) is fully relaxed, then the optimized structure is used to calculate physical properties, such as $P$ and band gap.

As shown in Fig. 2(a), the relaxed lattice constants ($a$, $b$, $c$) almost linearly increase with $U_{eff}$. Meanwhile, the calculated $P$ decreases from 15.5 μC/cm$^2$ down to 10.0 μC/cm$^2$, and the band gap $\Delta$ increases from 2.5 eV to 3.3 eV, as shown in Fig. 2(b). Taking the measured data on CTO single crystals as reference, $U_{eff}$=3 eV is the best choice to reproduce the experimental values: 1) the lattice constants $a$=5.475 Å, $b$=5.413 Å, $c$=19.393 Å, with the unit volume of 574.74 Å$^3$ (the experimental volume is 573.40 Å$^3$ [26]); 2) $P$=10.52 μC/cm$^2$ (the measured value: ~8.0 μC/cm$^2$ [10]); 3) the band gap $\Delta$~2.9 eV (the measured one ~3.6 eV [27]). Considering that the experimental polarization may be reduced a little bit due to unsaturation or other extrinsic factors, our estimated value gives a perfect description of real polarization, much improved than previous calculations. Furthermore, the 20% underestimation of band gap is acceptable considering the well-known methodological drawback of DFT.

The energy barrier between polar $A2_1am$ structure and nonpolar $Amam$ structure is 68.4 meV/Ti, close to (a little higher than) previous result 56 meV/Ti obtained using pure PBEsol without $U$ [13].

In short, the choice of PBEsol+$U$ ($U_{eff}$=3 eV) can give an improved description of structure and ferroelectricity of CTO, which will be adopted in following calculations of other compounds.

*3.2. Site-selective substituted CTO compounds*

As mentioned before, Mulder *et al*. [16] predicted a general rule to improve ferroelectric $P$ in the 327 RP series by substituting $A$' ion using larger ions. Two model systems Ca$_3$Sn$_2$O$_7$ and Ca$_3$Zr$_2$O$_7$ were tested. However, this general rule seems to contradict with the experimental observation in Sr-substituted CTO [10], in which $P$ is significantly suppressed by increasing Sr's concentration. We have also checked the results presented in Ref. [16], which can be successfully reproduced (as compared in Supplementary Materials [28]) despite the different software packages and pseudopotentials. Our following calculation will try to solve this puzzle.

First, as mentioned in the Introduction section, experimental works recommended the site-selective substitution, considering the different coordination numbers and spare space. However, considering the divergent results between theoretical prediction and experimental observation, it is necessary to perform a numerical check on this site-selective assumption, at least in the qualitative level.

The larger Sr and smaller Mg cations are expected to prefer the $A$' and $A$ sites, respectively. To verify this point, 1/3 Ca and 2/3 Ca are substituted by Sr and Mg respectively. Not only the aimed configurations of (CaTiO$_3$)$_2$SrO, (MgTiO$_3$)$_2$CaO, and (MgTiO$_3$)$_2$SrO are calculated, but also all other possibilities of configurations (i.e. Mg and Sr partially or fully occupy the *A'* and *A* sites respectively) are tested (see Fig. 3(a)). The relaxed CTO structure is adopted as the initial structure for the substituted cases.

As the first step, this CTO structure without further relaxation is used for a qualitative comparison. The calculated energy distributions are shown in Fig. 3(b), as a function of the number of anti-site pairs *p* (an anti-site pair means that one larger/smaller ion occupys the A/A' site). Our calculations indeed suggest that the lattices with Sr and Mg occupying *A'*-sites and *A*-sites are lower in energy comparing with those with anti-site occupancy. Interestingly, the energy with same anti-site occupancy ratio are almost identical, despite the details of configurations.

Next, these structures are further relaxed (without changing the space groups) for more accurate comparison, as summarized in Fig. 3c. Quantitatively, the average anti-site energies are reduced by about 65%, 75% and 85% for $Ca_2SrTi_2O_7$, $Mg_2CaTi_2O_7$, and $Mg_2SrTi_2O_7$, respectively. Even though, the qualitative conclusion for site preference remains unchanged. Considering the anti-site energy, the site-selective substitution is possible, even not ideal in real materials. In the following, these site-selective substituted compounds will be studied in details.

Second, starting from the $A2_1am$ structure, $(CaTiO_3)_2SrO$ is relaxed till the optimized structure. It remains $A2_1am$, and its calculated ferroelectric *P* is 10.46 μC/cm$^2$, almost identical to the CTO itself. The energy barrier between the polar $A2_1am$ structure and nonpolar *Amam* structure is greatly reduced to 5.7 meV/Ti, which is favorable to reduce the coercivity.

Using the same procedure, $(MgTiO_3)_2CaO$ and $(MgTiO_3)_2SrO$ are also calculated, as summarized in Table I. Their $A2_1am$ structures lead to 18.48 μC/cm$^2$ and 21.26 μC/cm$^2$, much improved comparing with that of CTO. The energy barriers between the polar $A2_1am$ structures and nonpolar *Amam* structures are 363.7 meV/Ti and 299.1 meV/Ti, much higher than that of CTO.

*3.3 Structural transition of substituted CTO compounds*

Till now, our DFT results on $(CaTiO_3)_2SrO$ remain inconsistent with the experimental observation. First, the suppressed *P* upon Sr-substitution observed in experiment has not been captured. Second, the experimental coercivity seems to be unchanged upon Sr-substitution [10]. These inconsistencies make the prediction on $(MgTiO_3)_2CaO$ and $(MgTiO_3)_2SrO$ becomes uncertain, although currently there is no experimental comparison available for these materials.

Recent two theoretical studies suggested other possible combinations of TiO$_6$ octahedral rotations/tiltings. The nonpolar *Pnam* structure was proposed to be the intermediate state during the ferroelectric switching, whose energy is only a little higher for 7 meV/Ti than the ground state [13]. In addition, the nonpolar *Pbcn* structure is also very close to the ground state, which can be stabilized by strain [14]. In this sense, it is very necessary to carefully re-check all possible space groups for the substituted CTO compounds. Here all (totally 13) space groups mentioned in Ref. [13] and [14] have been tested for CTO and its substituted CTO compounds. The results are summarized in Table. II.

For pure CTO, the polar *A2$_1$am* structure is indeed the lowest energy one. The proposed intermediate state *Pnam* during the ferroelectric switching is 7.7 meV/Ti higher, very close to previous result without *U* [13]. However, for (CaTiO$_3$)$_2$SrO, the real ground state is not *A2$_1$am* but *P4$_2$/mnm*, which is nonpolar. Thus, it becomes natural to understand the suppression of *P* in CTO upon Sr-substitution. In fact, the *P4$_2$/mnm* structure was indeed observed experimentally in a narrow window of Ca$_{3-x}$Sr$_x$Ti$_2$O$_7$ (0.9≤x≤1) [17].

In contrast, (MgTiO$_3$)$_2$CaO is similar to CTO. The *A2$_1$am* structure remains the ground state with almost doubled *P* (=18.48 μC/cm$^2$). Similar to the CTO case, the intermediate *Pnam* state remains the second lowest one, but the energy difference (15.5 meV/Ti higher) is also doubled comparing with CTO. According to the two-step switching path of CTO [13], the *P2$_1$am* stands for the energy barrier. Here for (MgTiO$_3$)$_2$CaO, the *P2$_1$am* is 37.1 meV/Ti higher than *A2$_1$am*, almost identical to that of CTO. Therefore, considering the doubled *P* of *A2$_1$am* structure, the coercive field should be even significantly reduced in (MgTiO$_3$)$_2$CaO, since the fundamental driving force for ferroelectric switching is *E·P* despite the switching paths. In other words, (MgTiO$_3$)$_2$CaO can be considered to be an enhanced CTO regarding its ferroelectricity.

For (MgTiO$_3$)$_2$SrO, the situation becomes quite different. The *P2$_1$nm* becomes the ground state which is also polar. The distortion modes are sketched in Fig. 4(d). The calculated *P* is 13.65 μC/cm$^2$, slightly higher than that of CTO. And the *P2$_1$am* is the second lowest energy structure, but the expected nonpolar intermediate *Pnam* state is very high in energy (~71.1 meV/Ti). Thus, the ferroelectric switching may become quite difficult in (MgTiO$_3$)$_2$SrO, with a very high coercive field, if there's no other route. Of course, the accurate swithcing pathes deserve further investigations.

The relaxed structures for these ground states can be found in the Supplementary Materials [28].

*3.4 Layer contribution in substituted CTO compounds*

As sketched in Fig. 1, $(A\text{TiO}_3)_2A'\text{O}$ can be divided into the $A$-O, Ti-O, and $A'$-O layers stacking along the $c$-axis. For a more intuitive understanding of the ferroelectric $P$ in these compounds, the charge dipole moment in each layer is calculated using the BEC model. The BEC's are calculated using DFPT for the four compounds in the ground state structure. More details of BEC values ($C$) are summarized in Supplementary Materials [28].

The evaluated BEC for each atom is similar in different compounds. However, at different sites, the identical element can show difference regarding the BEC values, all of which deviate from their nominal values more or less. The charge dipole of each layer $P_{\text{layer}}$ is estimated as [29]:

$$P_{\text{layer}} = \sum_i C_i \cdot \Delta r_i, \qquad (1)$$

where $i$ index ions in each layer; $\Delta r_i$ is the displacement vector of one ion from its high symmetric position in the $I4/mmm$ structure. $C_i$ is the tensor of BEC.

The total polarization can also be estimated using the BEC model, leading to 10.24 $\mu\text{C/cm}^2$, 16.85 $\mu\text{C/cm}^2$, and 13.23 $\mu\text{C/cm}^2$ for $\text{Ca}_3\text{Ti}_2\text{O}_7$, $(\text{MgTiO}_3)_2\text{CaO}$, and $(\text{MgTiO}_3)_2\text{SrO}$ respectively, which agrees with the values calculated by Berry phase method quite well. Thus the BEC model can be a reliable tool to analyze the contribution of $P$ from each layer.

The layer-by-layer dipole moments estimated from the BEC model are plotted in Fig. 5, for four compounds studied in this work. For $\text{Ca}_3\text{Ti}_2\text{O}_7$, the dipole moments are opposite between Ca(1)O and Ca(2)O layers, as revealed in previous studies. In details, Ca(1)O and TiO$_2$ layers contribute positively to the net $P$, while Ca(2)O contribute negatively. Then uncompensation between these dipoles lead to a net polarization along the $a$ axis. For $(\text{CaTiO}_3)_2\text{SrO}$, the $P4_2/mnm$ structure is highly symmetric, leading to absolute zero dipole moment in each layer. For $(\text{MgTiO}_3)_2\text{CaO}$, the contribution from MgO layer is much larger (~70%) than original one from Ca(1)O. The small size of Mg allows larger off-center displacement. The negative contribution from Ca(2)O layers is slightly enhanced (~10%). The TiO$_2$ layers become almost cento-symmetric. The final effect is the enhanced net $P$. For $(\text{MgTiO}_3)_2\text{SrO}$, the situation is rather different due to the $P2_1nm$ space group. All layers contribute positively to net $P$, while the main contribution is from SrO and TiO$_2$ layers.

**IV. Conclusion**

In summary, based on the PBEsol+$U$ method, we have corrected the over-estimation of polarization of $Ca_3Ti_2O_7$. The iso-valent substitution of two Ca sites in $Ca_3Ti_2O_7$ is studied. The site-selective substitution of Mg and Sr are qualitatively verified: Sr prefers the *A'* site while Mg prefers the *A* site. With this preference, the fully site-selective $(CaTiO_3)_2SrO$, $(MgTiO_3)_2CaO$, and $(MgTiO_3)_2SrO$ are calculated. Our calculations reveal that $(CaTiO_3)_2SrO$ prefers the *P4$_2$/mnm* non-polar structure rather the *A2$_1$am* polar phase, which can explain the experimentally observed suppression of ferroelectric polarization in $Ca_{3-x}Sr_xTi_2O_7$. In contrast, the polarization can be enhanced in $(MgTiO_3)_2CaO$, and $(MgTiO_3)_2SrO$. Especially for $(MgTiO_3)_2CaO$, the polarization is nearly doubled comparing with $Ca_3Ti_2O_7$, while its coercivity is almost unchanged. The $(MgTiO_3)_2SrO$ is a little more complex, which turns to own another polar space group. Our work suggests that the iso-valent substitution can significantly tune the hybrid improper ferroelectricity in the 327-type RP compounds.

**Acknowledgment**

This work was financially supported by the National Key Research Program of China (Grant Nos. 2016YFA0300101, 2015CB654602) the National Science Foundation of China (Grant Nos. 51431006, 51721001, 11674055).

**Figure Captions**:

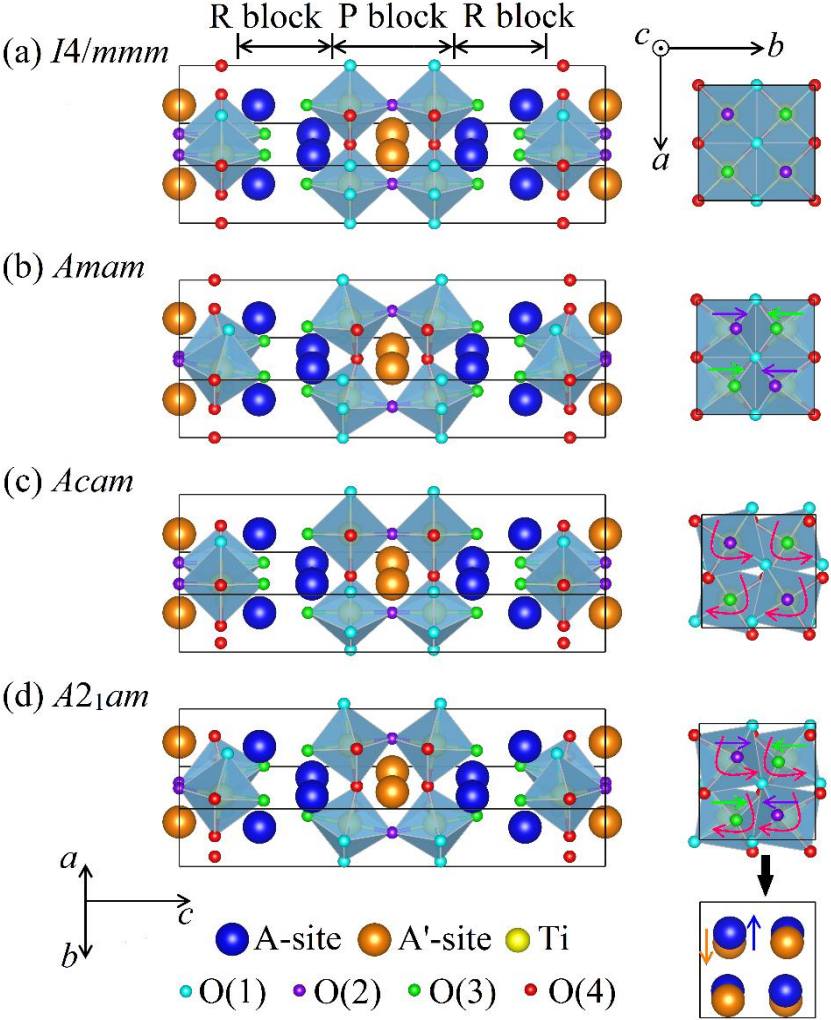

**Fig. 1.** The structures of 327-type RP titanates. (a) The parent structure *I4/mmm*, without octahedral rotation/titling. (b) The *Amam* structure, with octahedral tilting but no rotation. (c) The Acam structure, with octahedral rotation but no tilting. (d) The ground state of CTO, with both octahedral rotation and titling. The right panes are the corresponding top view from *c*-axis. The off-center displacements of *A*/*A'* ions are also shown.

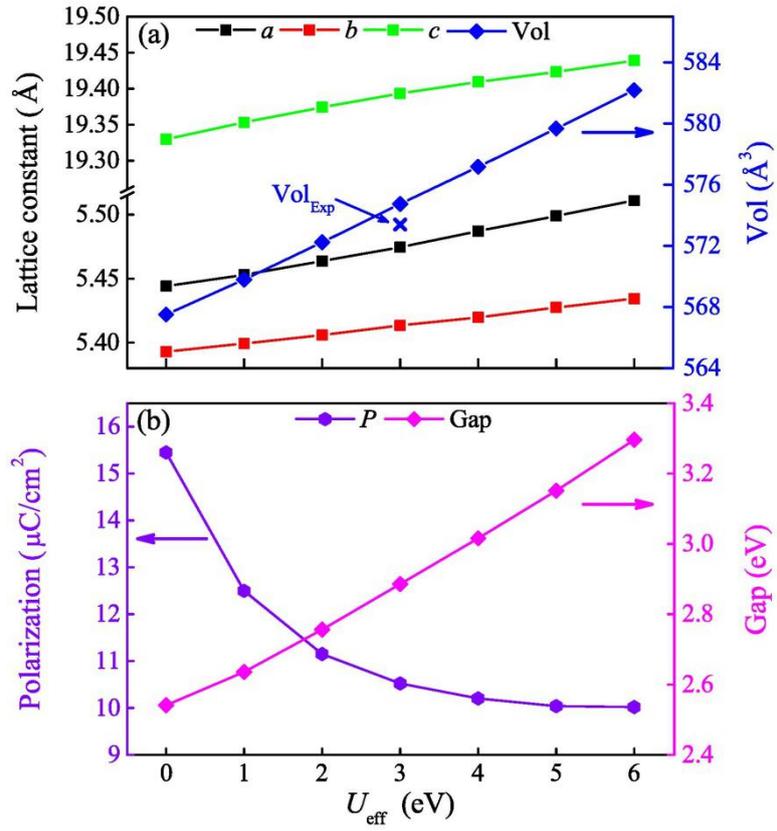

**Fig. 2.** The physical properties of CTO in our DFT calculation as a function of $U_{\text{eff}}$. (a) The lattice constants (left) and volume (right). (b) The ferroelectric $P$ (left) and the band gap (right).

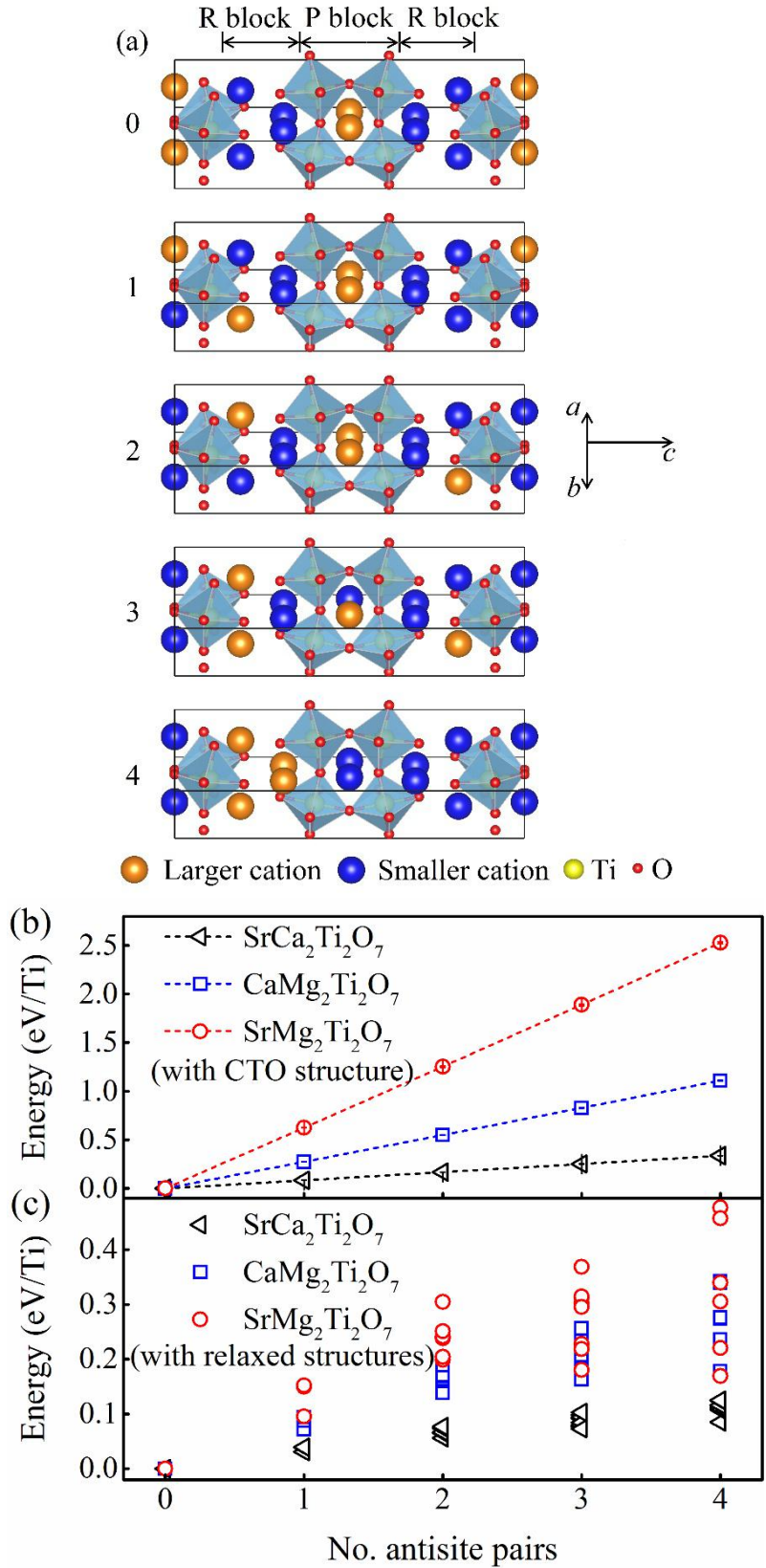

**Fig. 3.** (a) Different configuration of iso-valent substitution. The top one is the ideal limit with all larger/smaller ions occupying the *A'*/*A* sites. From top to bottom, more anti-site pairs are created (an anti-site pair means that one larger/smaller ion occupies the *A*/*A'* site). Here for each anti-site pair(s) number, only one structure is shown as an example, while there are more structures with the same number of anti-site pair(s). (b-c) Average energy per Ti as a function of the number of anti-site pairs. (b) Calculated using the optimized CTO structure without further relaxation. It should be noted that there are many configurations for each number of anti-site pairs. However, their energy are quite close to each other, characterized by the very small error bars. (c) Calculated starting from the optimized CTO structure with further relaxation. The multiple configurations for each number of anti-site pairs lead to divergent energy. But the configurations without any anti-site pairs remain the lowest energy ones.

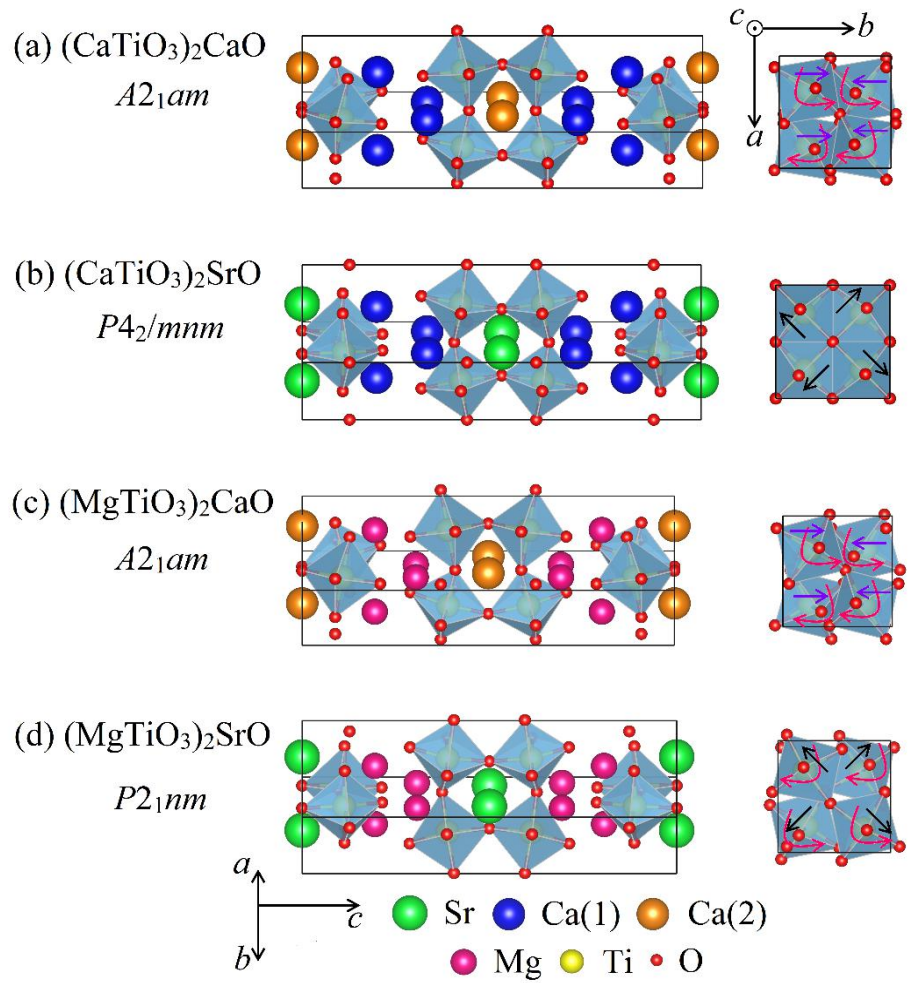

**Fig. 4.** The ground state structures of CTO and substituted CTO. (a) CTO. (b) (CaTiO$_3$)$_2$SrO, which becomes non-polar. (c) (MgTiO$_3$)$_2$CaO. (d) (MgTiO$_3$)$_2$SrO, which adopts another polar group.

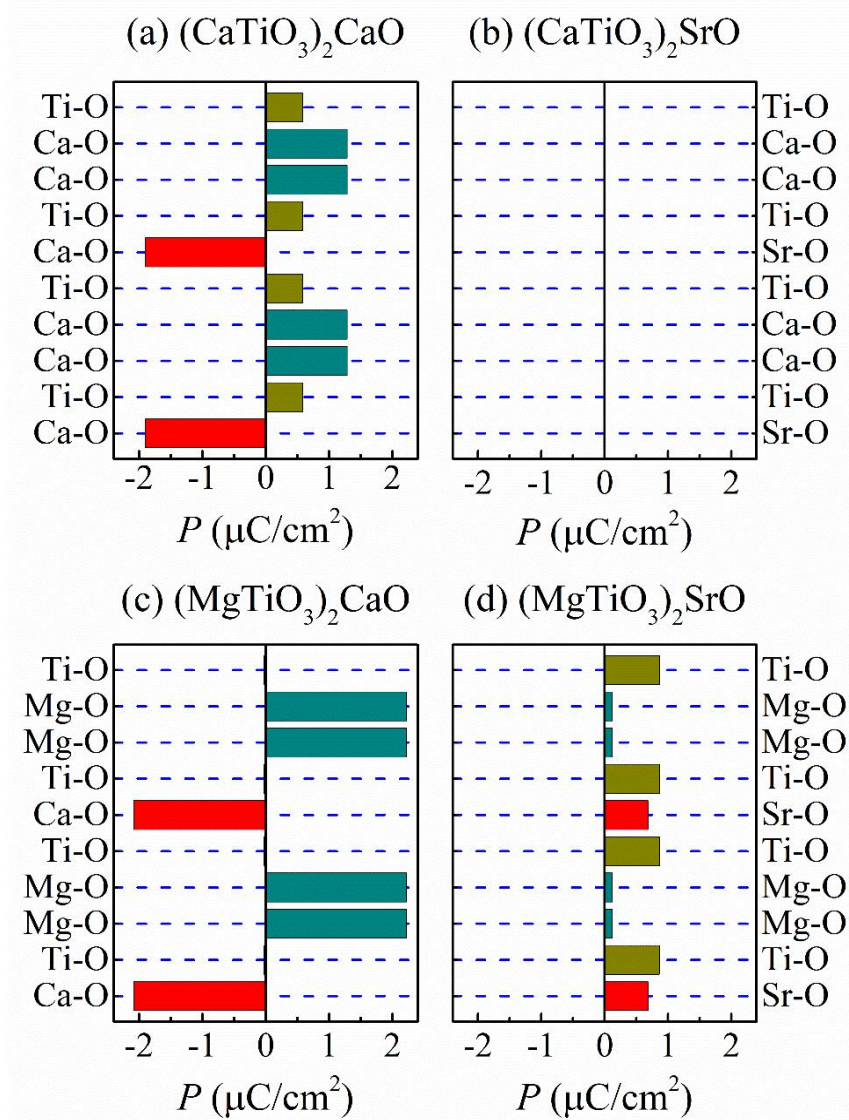

**Fig. 5.** The layer-by-layer contribution of polarization estimated from BEC model. (a) CTO. (b) (CaTiO$_3$)$_2$SrO. (c) (MgTiO$_3$)$_2$CaO. (d) (MgTiO$_3$)$_2$SrO.

**Table I.** Ferroelectric $P$ along the $a$ axis in the $A2_1am$ structure calculated using the Berry phase method. The corresponding energy differences $\Delta E$ between the $A2_1am$ phase and the $Amam$ structure are also shown.

|  | $(CaTiO_3)_2CaO$ | $(CaTiO_3)_2SrO$ | $(MgTiO_3)_2CaO$ | $(MgTiO_3)_2SrO$ |
| --- | --- | --- | --- | --- |
| $P$ (μC/cm$^2$) | 10.52 | 10.46 | 18.48 | 21.26 |
| $\Delta E$ (meV/Ti) | 68.4 | 5.7 | 363.7 | 299.1 |

**Table II.** The energies $E$ (meV/Ti) obtained from full structural relaxations with various space groups. The lowest-energy state for each compound is taken as the reference. Due to structural instability, some space groups become others after relaxation, and the output space groups are indicated. For each stable polar structure, the corresponding $P$ is also shown within parentheses, which is calculated using the Berry phase method.

| IT number | Group | $(CaTiO_3)_2CaO$ | $(CaTiO_3)_2SrO$ | $(MgTiO_3)_2CaO$ | $(MgTiO_3)_2SrO$ |
|---|---|---|---|---|---|
| 55 | $Pbam$ | $Acam$ | $Acam$ | $Acam$ | $Acam$ |
| 58 | $Pnnm$ | 59.8 | $P4_2/mnm$ | 270.7 | 183.4 |
| 60 | $Pbcn$ | 26.9 | 11.8 | 57.5 | 54.6 |
| 62 | $Pnam$ | 7.7 | 15.4 | 15.5 | 71.1 |
| 63 | $Amam$ | 68.4 | 18.3 | 363.7 | 343.0 |
| 64 | $Acam$ | 109.1 | 74.5 | 602.3 | 503.0 |
| 127 | $P4/mbm$ | 204.9 | 88.1 | 909.9 | 685.7 |
| 136 | $P4_2/mnm$ | 60.1 | 0 | 436.2 | 245.5 |
| 139 | $I4/mmm$ | 272.0 | 96.1 | 1176.8 | 842.0 |
| 6 | $Pm$ | $A2_1am$ | $P4_2/mnm$ | 53.2 (20.00 μC/cm$^2$) | 60.3 (17.75 μC/cm$^2$) |
| 26 | $P2_1am$ | $A2_1am$ | $Pnam$ | 37.1 (44.65 μC/cm$^2$) | 28.6 (43.93 μC/cm$^2$) |
| 31 | $P2_1nm$ | $A2_1am$ | $P4_2/mnm$ | 43.6 (12.47 μC/cm$^2$) | 0 (13.65 μC/cm$^2$) |
| 36 | $A2_1am$ | 0 (10.52 μC/cm$^2$) | 12.6 (10.46 μC/cm$^2$) | 0 (18.48 μC/cm$^2$) | 43.9 (21.26 μC/cm$^2$) |
| 38 | $C2mm$ | 52.3 (0.17 μC/cm$^2$) | $P4_2/mnm$ | $P2_1nm$ | $P2_1nm$ |